# Life history evolution and the origin of multicellularity: the case of different types of cells

Fuad Aleskerov[1]
Denis Tverskoy[2]

## Abstract

The problem of unicellular-multicellular transition is one of the main issues that is discussing in evolutionary biology. In [1] the fitness of a colony of cells is considered in terms of its two basic components, viability and fecundity. Intrinsic trade-off function of each cell defines a type of cell. We elaborate models providing in [1]. Assuming that all intrinsic trade-off functions are linear, we construct a model with different cell types and show that the differentiation of these types tends to full specialization. In addition, we attempt to consider the fact that environmental factors influence on the fitness of the colony. Thus, we introduce an energy restriction to the model and show that in optimum we get situations in which there exists a set of states, each of them allowing colony to achieve the same maximum level of fitness. In some states arbitrary chosen cell may be specialized, in some – unspecialized, but fecundity and viability of each cell belong to limited ranges (which are unique for each cell). It is worth pointing out that the models from [1] are not robust. We try to overcome this disadvantage.

## 1. Introduction

The problem of unicellular-multicellular transition is one of the main issues that is discussing in evolutionary biology. It is necessary to know how colonial organisms are transformed into multicellular organisms and what preconditions underlie this transition.

The separation of a body's tissues is the main characteristic of a multicellular organism. This separation means that the majority of cells in the organism is specialized for one function and loses the potential ability to specialize for other functions.

Some cells in a colonial organism may be specialized for specific functions but may not lose the ability to specialize elsewhere. If conditions that lead the colony to full specialization are performed over a long period of time, it is possible that the unicellular-multicellular transition will occur. Therefore, it is important to determine the conditions that contribute to the full specialization of a colonial organism. For example, it is reasonable to suspect that different cell types may cause full specialization. Another interesting question is the influence of environmental factors on the colony's behavior. We introduce these environmental factors into the model in the form of the total energy constraints consumed by considering colony of cells.

[1] alesk@hse.ru, International Laboratory for Decision Analysis and Choice and Department of Mathematics for Economics of the National Research University Higher School of Economics (NRU HSE), and Institute of Control Sciences of Russian Academy of Sciences

[2] disa1591@mail.ru, International Laboratory for Decision Analysis and Choice and Department of Mathematics for Economics of the National Research University Higher School of Economics (NRU HSE), and Institute of Control Sciences of Russian Academy of Sciences

Fundamental models investigated the problem of unicellular – multicellular transition were provided in [1]. These models best illustrate evolution of Volvocalean Green Algae but they may be applied to other lineages as well. In [1] the model has been illustrated on the example of volvocales green algae. These are flagellated photosynthetic organisms with coherent glycoprotein cell walls and represent the most appropriate system in research of the process of transition under study, because Volvocales linage ranges from the single-cell organisms to undifferentiated, soma-differentiated and germ-soma differentiated organisms arranged according to the size of the colony. Volvocales live in standing waters and so need flagellar beating in order to move toward light and nutrients. Therefore motility is an important factor contributing to viability of Volvocales [2]. Volvocales' type of cell division represents palintomy with multiple fission. Also, useful fact is that the species with increased cell specialization do not have a single origin.

However, the results in [1] look non-robust. Because of the identity of all cells in the colony assumed in [1], in optimum it does not matter which cells belong to the sets of soma-specialized cells or germ-specialized cells. So, if we change slightly some characteristics that not reflected explicitly in the model, sets of germ and soma-specialized cells changes – the model only requires that the ratio between their cardinality should remain constant. Thus, small changes in parameters may force soma-specialized cell become germ-specialized immediately. In linear case this non-robustness also lies in a fact that no more than the half cells can be soma-specialized and no more than the half cells can be germ-specialized. These facts have attracted our attention and we provide a new model to overcome above mentioned non-robustness of this beautiful model developed in [1].

**Structure of the text.**

We begin with the model, provided in [1] and describe it in Section 2. Then in Section 3 we propose the model with different cell types, extend this model taking into account energy constraints in Section 4. In section 5 we give a short survey of related works. Section 6 concludes.

**2. An overview of the models**

In [1], the authors constructed models that explore the fitness trade-offs at both the cell and group levels during the unicellular-multicellular transition. Thus, fitness is considered in terms of its two basic components: viability and fecundity. In [1], the trade-off function (1) is studied, which reflects the intrinsic relationships that link viability and fecundity within the cell due to cell physiology and other constraints.

Let v be viability and b represents fecundity. Then:

$$v = v_{max} - \alpha * b. \qquad (1)$$

In [1], the authors noted that in unicellular organisms, the cell must contribute to both of the fitness components. In multicellular groups, each cell may be unspecialized, such as in unicellular organisms, or, in contrast, may specialize only in the germ or only in the soma. This fact can lead to the formation of germ – soma ("G-S") specialization, in which some cells lose their autonomy in favor of the group and, as a result, their fitness and individuality are transferred from the cell level to the group level. In [1], the cases in which "G-S" specialization may occur are studied. It is noted as well that the models are most applicable to volvocine green algae.

There are two types of models that are considered in [1]: the fitness isocline model and the full optimization model. We discuss only the second model because this model is more general than the fitness isocline model. In the full optimization model, all of the cells are considered simultaneously, and the strategic purpose of the colony is to maximize its fitness. Below, we describe this model in detail to emphasize its advantages and disadvantages. Then we try to improve it.

### 2.1. Full optimization model

Consider a colony consisting of N cells, $i = 1 \ldots N$ are indices of cells in the colony, $b_i$ – resulting contribution of cell i to the fecundity of the group, $v_i$ – viability-enhancing capability of cell i.

The fitness trade-off function (continuous and determined on a convex hull) is more common than a linear function and should be the same for all of the cells from the colony:

$$v_i = v(b_i). \qquad (2)$$

The group's level of fecundity is an additive function of variable $b_i$. The group's level of viability is an additive function of variable $v_i$, i.e.,

$$B = \sum_{i=1}^{N} b_i; \quad V = \sum_{i=1}^{N} v_i. \qquad (3)$$

In [1], it is assumed that the group fitness, which we should maximize, is the product of the group viability and fecundity:

$$W = V * B. \qquad (4)$$

In [1], the problem of choosing the correct form of the group fitness function is also discussed. This type of function (4) is based on simple intuition. For instance, imagine that one cell in the group has a high level of fecundity but low viability, and another cell is strictly the opposite, with a high level of viability and low fecundity. Each of these cells by itself would have a low fitness, but together they can achieve a high fitness for the group. Function (4) considers this reasoning in contrast to, for example, the average fitness of all cells (5):

$$W = \frac{1}{N} \sum_{i=1}^{N} b_i * v_i \qquad (5)$$

However, we should note that most of the assumptions in [1] would still hold even if the fitness is described by a more general function with special properties.

In general, the full optimization model can be written formally as a type of optimization problem (6):

$$\begin{cases} W = \sum_{i=1}^{N} b_i * \sum_{i=1}^{N} v_i \to max_{b,v} \\ \forall i = 1..N: v_i = v(b_i), \\ b \geqq 0, \\ v \geqq 0. \end{cases} \quad (6)$$

Assume that there is no initial cost of reproduction in the model. In this case, the following results are true:

1. If the function $v(b)$ is strictly concave, then the group of cells should remain unspecialized.

2. If the function $v(b)$ is linear ($v = v_{max} - \alpha * b$), then the group of cells behaves as if there was just one cell; therefore, each cell is indifferent to specialization.

3. If the function $v(b)$ is strictly convex, then the group of cells aims for full specialization. In addition, if there is an even number of cells in the group, then half should specialize in the germ and a half in the soma. If there is an odd number of cells in the group, $\left[\frac{N}{2}\right]$ of these cells should specialize in the germ, $\left[\frac{N}{2}\right]$ should specialize in the soma and one cell should remain unspecialized.

An initial investment is necessary for reproduction. This investment requires an additional spending of energy, which can lead to the appearance of initial costs of reproduction that can be considered within the trade-off function. The initial costs of reproduction lead to full specialization in linear and convex cases of improved model and provide the opportunity for specialization in the concave case of this model.

Despite all of the fundamental results of the full optimization model, there are a wide variety of problems that cannot be solved using this model and some disadvantages that are connected strictly with biological processes that this model cannot describe. For instance,

1. The full optimization model does not explain why the cells are jointed into a group and exist in this state instead of continuing to exist separately.

2. In the full optimization model the separate form of fitness function of the colony is used. However, in some cases, this form is not appropriate.

3. Models, provided in [1] assumed that all of the cells in the colony should be identical. In some cases it is reasonable to omit this identity assumption.

4. The full optimization model does not strictly reflect the fact that some environmental factors may influence on the fitness of the colony. .

Below we try to construct a model that is based on two important assumptions:

1. Cells within the colony are of different types (different intrinsic trade-off functions).

2. There are some energy restrictions in the system under consideration.

In Section 2 the linear model, which satisfies only the first assumption, is studied. In Section 3, we construct the model that satisfies both of the specified assumptions.

## 3. Full optimization model with different cell types
### 3.1. Formulation of the problem

Now we will assume that a set of all cells in the colony is divided into finite number of non-intersected, non-empty subsets (we get a partition in the set of all cells), such that:

- Each two cells belonging to the same subset have the same intrinsic trade-off function (the same type);

- Each two cells from different subsets have different intrinsic trade-off functions (different types).

In order to simplify our analysis we suggest that all of the cells in the colony are unique, so each cell has its unique type. Note, that previous, general type of the model can be easily reduced to the model with unique cells in linear case.

Consider a group of N cells, $i = 1 \ldots N$ – indices of cells in the colony, $b_i$ - level of fecundity of cell i, $v_i$ — viability of cell i.

The group's level of fecundity is an additive function of variable $b_i$. The group's level of viability is an additive function of variable $v_i$:

$$B = \sum_{i=1}^{N} b_i; \quad V = \sum_{i=1}^{N} v_i. \tag{7}$$

We agree with [1] and continue to apply the type (8) of the group fitness functions because this function reflects the synergetic effects of jointly existing cells in the colony,

$$W = V * B. \tag{8}$$

The main extension of the model from [1] is that each cell has its own parameters of the trade-off function. Additionally, we assume that each individual's trade-off function is linear (9).

Thus,

$$\forall i = \overline{1, N}: v_i = v_i^{max} - \alpha_i * b_i, \tag{9}$$

$$\forall i = \overline{1, N}: \alpha_i > 0,$$

$$\forall i = \overline{1, N}: v_i^{max} > 0,$$

$$\forall i = \overline{1, N}: b_i^{max} = \frac{v_i^{max}}{\alpha_i}.$$

The main assumption of the full optimization model with different types of cells (strong differentiation of types) has following form:

$$\alpha_i \neq \alpha_j, \text{ for any } i, j \in \{1, .., N\}, i \neq j. \tag{10}$$

We use formula (10) because it presents the pure form of the differentiation of types in the model and because models with other types of differentiation can be easily transformed to this model.

Also, note that everywhere in this work, assumptions (9)-(10) are hold.

Therefore, we can construct the model as a classic optimization model with constraints,

$$\begin{cases} W = \sum_{i=1}^{N} b_i * \sum_{i=1}^{N} v_i \to max_{b,v} \\ \forall i = \overline{1, N}: v_i = v_i^{max} - \alpha_i * b_i, \\ \forall i = \overline{1, N}: b_i \geq 0, \\ \forall i = \overline{1, N}: v_i \geq 0. \end{cases} \tag{11}$$

This optimization problem can be transformed into a more convenient form (12):

$$\begin{cases} W = \sum_{i=1}^{N} b_i * \sum_{i=1}^{N} v_i \to max_{b,v} \\ \forall i = \overline{1, N}: v_i = v_i^{max} - \alpha_i * b_i, \\ \forall i = \overline{1, N}: b_i \geq 0, \\ \forall i = \overline{1, N}: v_i \geq 0. \end{cases} \leftrightarrow \begin{cases} W = \sum_{i=1}^{N} b_i * \sum_{i=1}^{N} (v_i^{max} - \alpha_i * b_i) \to max_b \\ \forall i = \overline{1, N}: 0 \leq b_i \leq b_i^{max}. \end{cases} \leftrightarrow$$

$$\leftrightarrow \begin{cases} W = \sum_{i=1}^{N} b_i * (\sum_{i=1}^{N} v_i^{max} - \sum_{i=1}^{N} \alpha_i * b_i) \to max_b \\ \forall i = \overline{1, N}: 0 \leq b_i \leq b_i^{max}. \end{cases} \tag{12}$$

Thus, we should maximize the polynomial of degree two, determined in the hyper parallelepiped in $R^n$.

### 3.2. An analysis of the model

In this subsection, we provide some propositions that show the nature and important properties of the solution to the optimization problem (12).

**Definition 1.** Let $\mathcal{P}$ be the class of optimization problems. $p \in \mathcal{P}$ if and only if p is an optimization problem with the form represented below:

$$\begin{cases} W_p = \left(\sum_{i \in I_1^p} b_i^{max} + \sum_{i \in I_3^p} b_i\right) * \left(\sum_{i \in I_2^p} v_i^{max} + \sum_{i \in I_3^p} v_i^{max} - \sum_{i \in I_3^p} \alpha_i * b_i\right) \to max_{b_i, i \in I_3^p} \\ \forall i \in I_3^p: 0 \leq b_i \leq b_i^{max}. \end{cases} \tag{13}$$

$$\text{where } \begin{cases} I_1^p \subset \{1,..,N\}; I_2^p \subset \{1,..,N\}; I_3^p \subseteq \{1,..,N\}, \\ I_1^p \cup I_2^p \cup I_3^p = \{1,..,N\}, \\ I_1^p \cap I_2^p = I_1^p \cap I_3^p = I_2^p \cap I_3^p = \emptyset, \\ |I_3^p| \geq 1. \end{cases} \quad (14)$$

Therefore, any task p from class $\mathcal{P}$ has a form (13) and is determined by a triplet of sets $(I_1^p, I_2^p, I_3^p)$ satisfying to the conditions (14). For example, problem (12), which is the formal description of the full optimization problem with different types of cells, also belongs to class $\mathcal{P}$ and is matched to the triplet $(\emptyset, \emptyset, \{1,..,N\})$ (denoted as PP).

Additionally, it is necessary to define some sets that will be useful below:

$$F_p - \text{ represents the domain of optimization task p}, p \in \mathcal{P},$$

$$H_p = \{b \in R^N: 0 \leq b_i \leq b_i^{max}, \forall i \in I_3^p; b_i = b_i^{max}, \forall i \in I_1^p; b_i = 0, \forall i \in I_2^p\}, p \in \mathcal{P},$$

$$\mathcal{H} = \{H_p, p \in \mathcal{P}\},$$

$$\mathcal{L} = \{H_p, |I_3^p| = 1\}.$$

$$F = \left\{ b \in H_{PP}: \exists l \in \mathcal{L}, \ b \text{ is a conditional stationary point of the problem} \begin{cases} W_p(b) \to max \\ b \in l \end{cases} \right\}.$$

**Theorem 1.** Let $b^*$ represents the solution to the problem (12). Then, the following statement is true:

$$(b^* \in Vert(H_{PP})) \vee (b^* \in F). \quad (15)$$

Where $Vert(H_{PP})$ represents a set of vertexes of $H_{PP}$.

The proof of Theorem 1 is provided in Appendix A.

Results of Theorem 1:

- If all types in the colony are unique, in the optimum there is no more than the one unspecialized cell. Furthermore, there cannot be a situation where all of the cells in a group are soma-specialized or germ-specialized. Also, this type of model is almost always robust.

- Consider more general case of this model, where we have a partition in the set of cells and each subset of cells from this partition has its unique type (suppose, all subsets from partition have cardinalities larger than one). According to Theorem 1, in the optimum some subsets from the partition consist from soma-specialized cells; other subsets – from germ-specialized cells; finally, the third type of subsets is either empty or contains only one subset of cells indifferent to specialization. In biological terms this case describes situations, where some cells in the colony are specialized and others established interlayer which is indifferent to specialization and may perform different functions depending on different external conditions. This type of model is almost always robust too.

Furthermore, models provided here allow to more than the half of the cells being specialized on soma (or germ) function.

Thus, we show that cell differentiation in linear cases tends the majority of cells to full specialization.

## 4. Full optimization model with an energy restriction and different types of cells
### 4.1. Formulation of the problem

All of the previous models have a serious disadvantage: environmental factors do not directly influence the fitness of a colony. Suppose that the colony has some way of obtaining energy from available resources. Let C be the level of the energy that is available to the colony by using a fixed way of obtaining energy and fixed environmental characteristics. It is necessary to have $k_1$ units of energy in order to construct a unit of fecundity and $k_2$ units of energy in order to support a unit of viability. Considering these judgments, we can construct an energy restriction as follows:

$$k_1 * \sum_{i=1}^{N} b_i + k_2 * \sum_{i=1}^{N} v_i \leq C, \text{ where } k_1 > 0, k_2 > 0, C > 0. \qquad (16)$$

Using this relationship, we obtain a full optimization model with an energy restriction and different types of cells.

$$\begin{cases} W = \sum_{i=1}^{N} b_i * \sum_{i=1}^{N} v_i \to max_{b,v} \\ \forall i = 1..N: v_i = v_i^{max} - \alpha_i * b_i, \\ k_1 * \sum_{i=1}^{N} b_i + k_2 * \sum_{i=1}^{N} v_i \leq C, \\ \forall i = 1..N: v_i \geq 0, \\ \forall i = 1..N: b_i \geq 0. \end{cases} \to \begin{cases} W(b) = \sum_{i=1}^{N} b_i * \sum_{i=1}^{N} (v_i^{max} - \alpha_i * b_i) \to max_b \\ k_1 * \sum_{i=1}^{N} b_i + k_2 * \sum_{i=1}^{N} (v_i^{max} - \alpha_i * b_i) \leq C; \\ \forall i = 1..N: 0 \leq b_i \leq b_i^{max}. \end{cases} \qquad (17)$$

The energy restriction can significantly influence on the behavior of the colony in the model. However, assume that this restriction allows colony to exist.

### 4.2. An analysis of the model

Below we provide some propositions that show the nature and important properties of the solution to the problem (17). The problem (17) is a task in which we should maximize a continuously differentiable function determined on a truncated hyper parallelepiped.

First of all, it is necessary to introduce sets that will be useful below.

$$M = \left\{ b \in R^N : k_1 * \sum_{i=1}^{N} b_i + k_2 * \sum_{i=1}^{N} (v_i^{max} - \alpha_i * b_i) = C \right\},$$

$$M_1 = \left\{ b \in R^N : k_1 * \sum_{i=1}^{N} b_i + k_2 * \sum_{i=1}^{N} (v_i^{max} - \alpha_i * b_i) \leq C \right\},$$

$$Q = F \cap M_1,$$

$$R = \left\{ b \in H_{PP} : \begin{cases} B(b) = \dfrac{C}{2k_1} \\ V(b) = \dfrac{C}{2k_2} \end{cases} ; \right\}.$$

**Theorem 2.**

Let $b^* \in R^N$ represents the solution to the problem (17). Then, the following statement is true:

$$(b^* \in Vert(H_{PP}) \cap M_1) \vee (b^* \in Q) \vee (b^* \in R) \vee [(b^* \in Vert(H_{PP} \cap M)) \wedge (R = \emptyset)] \tag{18}$$

Where $Vert(H_{PP})$ represents a set of vertexes of $H_{PP}$,

$Vert(H_{PP} \cap M)$ represents a set of vertexes of $H_{PP} \cap M$.

Generally speaking, this theorem determines the location of the optimal point and claims that there can be only three cases, each of them can be realized depending on the values of parameters of the model:

1. All cells in the colony are specialized: some in soma, some in germ;

2. There is the one unspecialized cell in the colony;

3. There exist a set of states, each of them allows colony to achieve the same maximum level of fitness. In some states the cell may be specialized, in some – unspecialized, but fecundity and viability of each cell belong to limited ranges (which are unique for each cell).

We can point out as well that the result of the full optimization model with different types of cells is robust. Indeed, small variation of parameters, which are not reflected explicitly in the model, cannot lead to sharp changes in the solution. The solution either does not change (as in the cases 1 and 2) or change slightly (as in the case 3 due to the fact that the optimal set of states is connected). Also, more than half of the cells may specialize in soma or in germ.

The proof of Theorem 2 is given in Appendix B.

According to the Theorem 2, in some cases we can get an energetic estimation of the optimal value of fitness function:

Let $b^* \in R^N$ represents the solution to the problem (17). Suppose $R \neq \emptyset$. Then, $W(b^*) \geq \dfrac{C^2}{4*k_1*k_2}$, meaning that under some circumstances, if the colony has the available level C of the energy, this energy can be transformed into no less than $\dfrac{C^2}{4*k_1*k_2}$ units of fitness.

## 5. A survey of the literature

There are a lot of papers dealing with the problem of unicellular–multicellular transition in terms of fitness and its two basic components – fecundity and viability. Here we provide short descriptions of some articles which characterize precisely main developments in discussed issues.

In [2] the problem of transition from unicellular to multicellular organism is studied using some physical assumptions concerning different processes in the organism in terms of physical laws, for example, hydrodynamic laws.

In [3] a general mathematical model about the division of labor is introduced. The main assumption of the model is that modules can contribute to two different tasks, which connected by a trade-off. It is shown that three factors favor that division of labor – positional effects, accelerating performance functions and interaction between modules.

In [4] the models of social choice have been applied to the problem under study. It has been shown that applying axiomatic approach allows constructing different social welfare functions describing the types of fitness-ranking on the set of alternatives, representing all states of the world that are relevant for the group under study. Authors suggest to apply extensive social welfare functions and their axiomatic to describe the fitness-functions of colonies and show that this axiomatic does not contradict the fact that the transition from unicellular to multicellular organism accompanies to replacing concavity by convexity in trade-offs.

In [5] a diversity in Volvocalean green algae is investigated in terms of the process of transition to multicellularity. Authors show that costs of reproduction plays an important role in the evolution of multicellularity in Volvocales. The proposed model allows explaining the GS – GS/S – G/S form of process of transition to multicellularity.

## 6. Conclusion

We have explored the fitness trade-offs between the basic components of survival and reproduction at both the cell and group levels during the unicellular–multicellular transition. We have considered the model that describes a colony with different types of cells and linear intrinsic trade-off functions.

As a result, if all types in the colony are unique, in the optimum there is no more than the one unspecialized cell. If we have a partition in the set of cells, and each subset of cells from this partition has its unique type, then in the optimum one part of subsets from partition consists from soma-specialized cells; second part – from germ-specialized cells; the third part is either empty or contains only one subset of cells indifferent to specialization. In biological terms this case describes situations, where some cells in the colony are specialized and others established interlayer which is indifferent to specialization and may perform different functions depending on different external conditions. Also, provided models are almost always robust. All in all, we show that cell differentiation in linear cases tends the majority of cells to full specialization, while in linear cases in [1] the cells are indifferent to specialization.

Additionally, we add environmental factors which directly influence the fitness of a colony. So, we have introduced a power restriction to the model and explored the full optimization model with a power restriction and different types of cells. As a result, in optimum we get situations in which there exists a set of states, each of them allows colony to achieve the same maximum level of fitness. In some states arbitrary chosen cell may be specialized, in some – unspecialized, but fecundity and viability of each cell belong to limited ranges (which are unique for each cell).

We can point out as well that the solution to the full optimization model with different types of cells and energy constraint is robust.

**Appendix A.**

*Introduce following notation:*

$Int\ A$ – an interior of a set A,

$ReInt\ A$ – a relative interior of a set A,

$\partial A$ - a boundary of a set A,

$Vert A$ - a set of all vertexes of a convex set A,

$\nabla f$ - gradient of a function f.

**Proof of Theorem 1.** First, let us proof the following

**Lemma 1.** For each $p \in \mathcal{P}, |I_3^p| > 1$, let $b_p \in Int F_p$; then, $b_p$ cannot be a solution to optimization problem p.

**Proof.** Choose any $p \in \mathcal{P}, |I_3^p| > 1$. Let $b_p \in Int F_p$. Suppose $b_p$ is a solution to optimization problem p.

1. $W_p(b_p) \geq W_p(b), \forall b \in F_p$, because $b_p$ is a solution to optimization problem p.

2. $\exists U_\varepsilon(b_p) \subset F_p : W_p(b_p) \geq W_p(b), \forall b \in U_\varepsilon(b_p)$, because $b_p$ is a solution to optimization problems p and $b_p \in Int F_p$. Therefore, $b_p$ is a local extremum point of function $W_p$.

3. Because $W_p$ is a continuously differentiable function on the entire domain, the necessary condition for a local extremum of the function $W_p$ at $b_p$ has the form: $\nabla W_p(b_p) = 0$.

$$\nabla W_p(b_p) = 0 \leftrightarrow$$

$$\frac{\partial W}{\partial b_j}(b_p) = \left( \sum_{i \in I_2^p} v_i^{max} + \sum_{i \in I_3^p} v_i^{max} - \sum_{i \in I_3^p} \alpha_i * b_i^p \right) - \alpha_j * \left( \sum_{i \in I_1^p} b_i^{max} + \sum_{i \in I_3^p} b_i \right) = 0, \forall j \in I_3^p;$$

$$\sum_{i \in I_3^p}(\alpha_i + \alpha_j) * b_i^p = \left(\sum_{i \in I_2^p} v_i^{max} + \sum_{i \in I_3^p} v_i^{max} - \alpha_j * \sum_{i \in I_1^p} b_i^{max}\right), \forall j \in I_3^p;$$

We obtain the system of $|I_3^p|$ linear equations. Since $|I_3^p| > 1$, we can choose two different linear equations with indexes $j$ and $k$ accordingly:

$$\begin{cases} \sum_{i \in I_3^p}(\alpha_i + \alpha_j) * b_i^p = \left(\sum_{i \in I_2^p} v_i^{max} + \sum_{i \in I_3^p} v_i^{max} - \alpha_j * \sum_{i \in I_1^p} b_i^{max}\right) \\ \sum_{i \in I_3^p}(\alpha_i + \alpha_k) * b_i^p = \left(\sum_{i \in I_2^p} v_i^{max} + \sum_{i \in I_3^p} v_i^{max} - \alpha_k * \sum_{i \in I_1^p} b_i^{max}\right) \end{cases}$$

Subtract the second equation from the first equation. Since

$$\sum_{i \in I_3^p}(\alpha_j - \alpha_k) * b_i^p = -(\alpha_j - \alpha_k) * \sum_{i \in I_1^p} b_i^{max}$$

then

$$(\alpha_j - \alpha_k) * \left(\sum_{i \in I_3^p} b_i^p + \sum_{i \in I_1^p} b_i^{max}\right) = 0$$

$$\sum_{i \in I_3^p} b_i^p = -\sum_{i \in I_1^p} b_i^{max} < 0 \to \exists i \in I_3^p : b_i^p < 0 \to b_p \notin F_p$$

We have a contradiction; therefore, $b_p$ cannot be a solution to optimization problem p. Lemma is proved.

Now we prove Theorem 1. Consider the optimization problem (12).

1. If N=1, then, obviously, the statement (15) is true.

2. Suppose N>1. Then, optimization problem (12) belongs to the class $\mathcal{P}$. Consider a point $b \in \text{Int}(H_{PP})$, (note, that $H_{PP} = F_{PP}$). According to Lemma 1, b cannot be a solution to optimization problem (12); therefore, the solution to (12) belongs to the set $\partial H_{PP}$.

Consider the behavior of function W in $\partial H_{PP}$. Choose one variable and prescribe it the value 0 or $b_i^{max}$. Thus, we describe a set $\partial H_{PP} = \bigcup_{p \in \mathcal{P}: |I_3^p| = N-1} H_P$. Therefore, we obtain tasks, each of which would be equivalent to some $p \in \mathcal{P}: |I_3^p| = N - 1$. According to Lemma 1, for each task $p \in \mathcal{P}: |I_3^p| = N - 1$, for all $b \in \text{ReInt}(H_P)$, it follows that b is not a solution to the problem (12). Therefore, we should find the solution to (12) in the set $\bigcup_{p \in \mathcal{P}: |I_3^p| = N-2} H_P$ where our conclusions are similar. We should repeat our procedure until we obtain tasks in the edges and vertexes of $H_{PP}$. The solution to (12) belongs to the vertexes or edges of hyper parallelepiped (12). Therefore:

$$\left(b^* \in \text{Vert}(H_{PP})\right) \vee (b^* \in F),$$

So, Theorem 1 is proved.

### Appendix B.

**Proof of Theorem 2.** First, we state the lemma.

**Lemma 2.** Consider optimization problem (19):

$$\begin{cases} W(b) = \sum_{i=1}^{N} b_i * \sum_{i=1}^{N} (v_i^{max} - \alpha_i * b_i) \to max_b \\ k_1 * \sum_{i=1}^{N} b_i + k_2 * \sum_{i=1}^{N} (v_i^{max} - \alpha_i * b_i) = C, \\ \forall i = 1..N: 0 \leq b_i \leq b_i^{max}. \end{cases} \quad (19)$$

Let $b^{**} \in R^N$ be the solution to problem (19). Then, the following statement is true:

$$[b^{**} \in R] \vee [(b^{**} \in Vert(H_{PP} \cap M)) \wedge (R = \emptyset)] \quad (20)$$

**Proof.** First, look at the class of optimization problems $\mathcal{U}$, which is constructed based on the coefficients of the task (19). Each task $u(I_1^u, I_2^u, I_3^u) \in \mathcal{U}$ has a type (21) (where $I_1^u \subseteq \{1,..,N\}, I_2^u \subset \{1,..,N\}, I_3^u \subset \{1,..,N\}$ : $I_1^p \cup I_2^p \cup I_3^p = \{1,...,N\}; I_1^p \cap I_2^p = I_1^p \cap I_3^p = I_2^p \cap I_3^p = \emptyset; |I_3^u| \geq 2$):

$$\begin{cases} W_u(b) = \left(\sum_{i \in I_1^u} b_i^{max} + \sum_{i \in I_3^u} b_i\right) * \left(\sum_{i \in I_2^u} v_i^{max} + \sum_{i \in I_3^u} v_i^{max} - \sum_{i \in I_3^u} \alpha_i * b_i\right) \to max_{b_i, i \in I_3^u} & (21.1) \\ k_1 * \left(\sum_{i \in I_1^u} b_i^{max} + \sum_{i \in I_3^u} b_i\right) + k_2 * \left(\sum_{i \in I_2^u} v_i^{max} + \sum_{i \in I_3^u} v_i^{max} - \sum_{i \in I_3^u} \alpha_i * b_i\right) = C & (21.2) \end{cases}$$

Each $u(I_1^u, I_2^u, I_3^u) \in \mathcal{U}$ represents a task of maximizing the function of the appropriate form (21.1) on the hyper plane of the appropriate form (21.2) in $R^{|I_3^u|}$ space. We choose and fix some $u(I_1^u, I_2^u, I_3^u) = u(I_1, I_2, I_3) \in \mathcal{U}$ and simplify the form of optimization problem u.

$$\begin{cases} B(b_i | i \in I_3) * V(b_i | i \in I_3) \to max_{b_i, i \in I_3} & (22.1) \\ k_1 * B(b_i | i \in I_3) + k_2 * V(b_i | i \in I_3) = C & (22.2) \end{cases} \quad (22)$$

$$\begin{cases} w(B(b_i | i \in I_3)) = \frac{1}{k_2} * B(b_i | i \in I_3) * (C - k_1 * B(b_i | i \in I_3)) \to max_{b_i, i \in I_3} \\ k_1 * B(b_i | i \in I_3) + k_2 * V(b_i | i \in I_3) = C \end{cases},$$

$$\text{where:} \begin{cases} B(b_i | i \in I_3) = \left( \sum_{i \in I_1} b_i^{max} + \sum_{i \in I_3} b_i \right) \\ V(b_i | i \in I_3) = \left( \sum_{i \in I_2} v_i^{max} + \sum_{i \in I_3} v_i^{max} - \sum_{i \in I_3} \alpha_i * b_i \right) \end{cases}$$

This task is related to the problem below:

$$f(b_i | i \in I_3) = w\big(B(b_i | i \in I_3)\big) = \frac{1}{k_2} B(b_i | i \in I_3) * \big(C - k_1 * B(b_i | i \in I_3)\big) \to max_{b_i, i \in I_3} \quad (23)$$

The necessary condition for a local extremum of the function $f$ has the form:

$$\frac{\partial w\big(B(b_1; \ldots; b_{|I_3|})\big)}{\partial b_i} = \frac{\partial w(B)}{\partial B} * \frac{\partial B(b_i | i \in I_3)}{\partial b_i} = \frac{\partial w(B)}{\partial B} = 0; \forall i \in I_3$$

This system is equivalent to one equation:

$$\frac{\partial w(B)}{\partial B} = 0 \to B = \frac{C}{2 * k_1} \quad (24)$$

Moreover, it is important to note that the function $f(b_i | i \in I_3): R^{|I_3|} \to R$ is concave in $R^{|I_3|}$, because:

1. The function $w(B) = \frac{1}{k_2} B * (C - k_1 * B)$ is concave in R.

2. The function $B(b_1; \ldots; b_{|I_3|}) = \left( \sum_{i \in I_1} b_i^{max} + \sum_{i \in I_3} b_i \right)$ is linear in its domain.

Suppose $x \in R^{|I_3|}, y \in R^{|I_3|}$. Then:

$$f(\lambda * x + (1 - \lambda) * y) = w(B(\lambda * x + (1 - \lambda) * y)) = w(\lambda * B(x) + (1 - \lambda) * B(y))$$
$$\geq \lambda * w\big(B(x)\big) + (1 - \lambda) * w\big(B(y)\big) = \lambda * f(x) + (1 - \lambda) * f(y), \forall \lambda \in [0; 1]$$

These equalities indicate that each stationary point of the function $f(b_1; \ldots; b_{|I_3|})$ in $R^{|I_3|}$ space represents the point of a local and a global maximum of the function $f(b_1; \ldots; b_{|I_3|})$. Moreover, all of the points of the global maximum of $f$ belong to a ($|I_3|$-1)-dimensional hyper plane, which is described by the equation below:

$$\sum_{i \in I_3} b_i = \frac{C}{2 * k_1} - \sum_{i \in I_1} b_i^{max} \quad (25)$$

Task (22) is equivalent to task (23) with restriction (22.2). This restriction represents a ($|I_3| - 1$)-dimensional hyper plane as well. As a result, all of the points that belong to the intersection of planes (22.2) and (25) are solutions to a problem of maximizing function (22.1) with restriction (22.2).

Therefore, the solutions to problem (22) satisfy the system of two linear equations:

$$\begin{cases} \sum_{i \in I_3} b_i = \frac{C}{2 * k_1} - \sum_{i \in I_1} b_i^{max} \\ \sum_{i \in I_3} (k_1 - k_2 * \alpha_i) b_i = C - k_1 * \sum_{i \in I_1} b_i^{max} - k_2 * \left( \sum_{i \in I_2} v_i^{max} + \sum_{i \in I_3} v_i^{max} \right) \end{cases} \quad (26)$$

These two planes are parallel or represent one plane when their normal vectors are linear-dependent vectors:

$$\begin{pmatrix} k_1 - k_2 * \alpha_1 \\ \dots \\ k_1 - k_2 * \alpha_{|I_3|} \end{pmatrix} = a * \begin{pmatrix} 1 \\ 1 \\ 1 \end{pmatrix} \rightarrow k_1 - k_2 * \alpha_1 = k_1 - k_2 * \alpha_{|I_3|} \rightarrow \alpha_1 = \alpha_{|I_3|}$$

According to assumption (10) about values of parameters of the model, $\alpha_1 \neq \alpha_{|I_3|}$. As a result, planes under consideration have a nonempty intersection; moreover, solutions to problem (22) belong to a $(|I_3| - 2)$-dimensional plane in $R^{|I_3|}$ space that can be described using system (26). Additionally, it is important to mention the statement below:

Let b satisfy (26). Then:

$$W(b) > W(b_1), \forall b_1 \in M, \text{ but not satisfying (26).} \quad (27)$$

This proposition follows from the curvature of optimized function and its concavity.

Note that all of the statements considered earlier in this lemma for the fixed optimization problem $u(I_1, I_2, I_3)$ remain valid for each $u(I_1^u, I_2^u, I_3^u) \in \mathcal{U}$.

Now we have a useful approach to investigate the problem (19). We emphasize the fact that the problem (19) represents the task $\tilde{u}(\emptyset, \emptyset, \{1, \dots, N\}) \in \mathcal{U}$ with additional restrictions $0 \leq b_i \leq b_i^{max}, \forall i = 1..N$. As found earlier, all of the solutions to task $\tilde{u}$ belong to a (N-2)-dimensional plane, that is why there are two possible cases:

1. There exists at least one point that simultaneously belongs to the optimal hyper plane and to the hyper parallelepiped $H_{PP} = \{b \in R^N : 0 \leq b_i \leq b_i^{max}, \forall i = 1..N\}$. According to condition (27), these and only these points represent the solutions to (19).

2. There is no point that simultaneously belongs to the optimal hyper plane (26) and to the hyper parallelepiped $H_{PP}$. According to condition (27) and due to the concavity of optimized function in energy restriction, we should search for the solution to the task (19) in $\partial H_{PP} \cap M$. We also obtain tasks $u(I_1^u, I_2^u, I_3^u) \in \mathcal{U}$, which are determined in appropriate hyper parallelepipeds. Because initial system (26) has no solution, corresponding systems (26) constructed for reduced tasks would have no solutions. We can repeat this procedure until we receive vertices of the set $H_{PP} \cap M$.

Therefore, we show that the following statement is true:

$$[b^{**} \in R] \vee \left[ (b^{**} \in Vert(H_{PP} \cap M)) \wedge (R = \emptyset) \right].$$

So, Lemma 2 is proved. According to Theorem 1 and Lemma 2 it is obvious that the following statement is true:

$$(b^* \in Vert(H_{PP}) \cap M_1) \vee (b^* \in Q) \vee [b^* \in R] \vee \left[\left(b^* \in Vert(H_{PP} \cap M)\right) \wedge (R = \emptyset)\right].$$

Theorem 2 is proved.

**Acknowledgments**

Fuad Aleskerov thanks Professor John A. Weymark who attracted his attention to the problem.

Authors thank DeCAn laboratory of NRU HSE for partial financial support.